\documentclass[conference]{IEEEtran}
\IEEEoverridecommandlockouts
\usepackage{cite}
\usepackage{amsmath,amssymb,amsfonts}
\usepackage{algorithmic}
\usepackage[graphicx]{realboxes}

\usepackage{multirow}
\usepackage{adjustbox}
\usepackage{tabularx}
\usepackage{amsmath}
\usepackage{amssymb}
\usepackage{adjustbox}

\usepackage{graphicx}
\usepackage{textcomp}
\usepackage{xcolor}
\def\BibTeX{{\rm B\kern-.05em{\sc i\kern-.025em b}\kern-.08em
    T\kern-.1667em\lower.7ex\hbox{E}\kern-.125emX}}
\begin{document}

\title{Decoding Visual Recognition of Objects from EEG Signals based on Attention-Driven Convolutional Neural Network\\}
\author{\IEEEauthorblockN{Jenifer Kalafatovich}
\IEEEauthorblockA{\textit{Dept. Brain and Cognitive Engineering} \\
\textit{Korea University}\\
Seoul, Republic of Korea \\
jenifer@korea.ac.kr}
\and
\IEEEauthorblockN{Minji Lee}
\IEEEauthorblockA{\textit{Dept. Brain and Cognitive Engineering} \\
\textit{Korea University}\\
Seoul, Republic of Korea \\
minjilee@korea.ac.kr}
\and
\IEEEauthorblockN{Seong-Whan Lee}
\IEEEauthorblockA{\textit{Dept. Artificial Intelligence} \\
\IEEEauthorblockA{\textit{Dept. Brain and Cognitive Engineering}}
\textit{Korea University}\\
Seoul, Republic of Korea \\
sw.lee@korea.ac.kr}

\thanks{This work was supported in part by the Institute for Information \& Communications Technology Promotion (IITP) grant, funded by the Korea government (MSIT) (No. 2015-0-00185, Development of Intelligent Pattern Recognition Softwares for Ambulatory Brain Computer Interface; No. 2017-0-00451, Development of BCI based Brain and Cognitive Computing Technology for Recognizing User's Intentions using Deep Learning; No. 2019-0-00079, Artificial Intelligence Graduate School Program (Korea University)).}

\thanks{© 20xx IEEE. Personal use of this material is permitted. Permission from IEEE must be obtained for all other uses, in any current or future media, including reprinting/republishing this material for advertising or promotional purposes, creating new collective works, for resale or redistribution to servers or lists, or reuse of any copyrighted component of this work in other works.}
}
\maketitle

\begin{abstract}
The ability to perceive and recognize objects is fundamental for the interaction with the external environment. Studies that investigate them and their relationship with brain activity changes have been increasing due to the possible application in an intuitive brain-machine interface (BMI). In addition, the distinctive patterns when presenting different visual stimuli that make data differentiable enough to be classified have been studied. However, reported classification accuracy still low or employed techniques for obtaining brain signals are impractical to use in real environments. In this study, we aim to decode electroencephalography (EEG) signals depending on the provided visual stimulus. Subjects were presented with 72 photographs belonging to 6 different semantic categories. We classified 6 categories and 72 exemplars according to visual stimuli using EEG signals. In order to achieve a high classification accuracy, we proposed an attention driven convolutional neural network and compared our results with conventional methods used for classifying EEG signals. We reported an accuracy of 50.37 $\pm$ 6.56\% and 26.75 $\pm$ 10.38\% for 6-class and 72-class, respectively. These results statistically outperformed other conventional methods. This was possible because of the application of the attention network using human visual pathways. Our findings showed that EEG signals are possible to differentiate when subjects are presented with visual stimulus of different semantic categories and at an exemplar-level with a high classification accuracy; this demonstrates its viability to be applied it in a real-world BMI.\\
\end{abstract}

\begin{IEEEkeywords}
electroencephalography (EEG), visual recognition, attention, convolutional neural network (CNN), brain-machine interface
\end{IEEEkeywords}

\section{Introduction}
Object perception and recognition are two of the most important abilities for humans; it allows us the interaction with the external environment and with other people~\cite{kaneshiro}. Humans have evolved to be able to rapidly identify objects and their semantic category when presented either visually or aurally. Specifically, the brain can extract such information from a visual stimulus in less than 200 msec depending on the semantic category of the presented stimulus~\cite{cichy}. Previous studies have investigated brain activities and mechanisms related to the process of transforming perceptual information into meaningful information~\cite{kaneshiro}. Visual information is processed in the occipital area of the brain; additionally, the two-stream hypothesis of visual processing states that the feature related to objects and spatial relationships among objects are encoded in the ventral-temporal stream and the dorsal-parietal stream (what and where pathways), respectively \cite{goodale}. It has been found that according to the semantic category of presented images different brain regions of the mentioned areas are activated preferentially \cite{hanson, haxby}. A new trend in brain-machine interface (BMI) is visual imagery, in contrast to visual perception, visual imagery can be defined as the mental representations accompanying with a perceptual experience of a certain stimulus\cite{pearson}. Previous studies have found changes in alpha band over occipital areas related to visual imagery\cite{kosmyna}. 

Electroencephalography (EEG), magnetoencephalography (MEG), and functional magnetic resonance imaging (fMRI) techniques have been used to explore brain activity related to the presentation of a stimulus of a certain semantic category \cite{Das, wang}. Haxby et al.~\cite{haxby} recorded fMRI signals while subjects were presented with a stimulus from different semantic categories (human faces, cats, houses chairs, scissors, shoes, bottles, and nonsense images). Data was analyzed to find evoked patterns in the ventral object vision pathway. The similarity between patterns was analyzed using correlations of brain responses. However, the use of fMRI techniques is not feasible for daily life applications. On the other hand, EEG provides more practical techniques due to its size and low cost, therefore it has been used in multiple applications to interpret brain signals such as evoked potentials ~\cite{Kwak,kim2}, event-related potentials (ERP)~\cite{Chen,minho}, motor imagery (MI)~\cite{suk} and sleep monitoring~\cite{lee2}.

One disadvantage when analyzing EEG signals is its high noise to signal rate~\cite{Lee}, therefore it is necessary to extract relevant features for a better analysis. Previous studies reported changes in the ERP waveforms when comparing different semantic categories signals, especially when comparing faces vs. objects~\cite{Itier}. Significant negative activity was found around 120 and 200 msec after stimulus onset depending on the stimulus category~\cite{Bentin, Ganis}. Carlson et al.~\cite{Carlson} used MEG to investigate the brain activity patterns related to encoding object category information. Subjects were presented with 24 object exemplars that belong to a certain category. Principal component analysis was used to reduce the dimensionality of data and objects were classified using linear discriminant analysis (LDA) (classification was done considering pairwise classes); performance was compared using different time windows. The difference between inanimate and animate objects was determined after 240 msec. Another study used a convolutional neural network (CNN) to classify targets and non-targets images. Subjects were presented with images of faces and cars (target and non-target respectively) during 500 msec; binary classification was performed obtaining 81.60 $\pm$ 0.07\% as average classification accuracy~\cite{Cecotti}. 

CNN has been applied in many areas such as classification for image, text, even brain signals, but has not been able to achieve the best performance yet~\cite{Kwak}. Recently, attention networks have been used~\cite{Wang2} in an attempt to improve performance. In vision recognition, this method uses a mask that modifies the input of the network and changes the importance of certain objects or regions of the image~\cite{Li}. Attention models have given successful results when applied to image and text classification~\cite{Fu}. Even in brain-computer interface, it has been used to improve the classification performance~\cite{Tan}. This method has shown promising results in increasing classification performance~\cite{Fu}. 

In this study, we attempt to decode the semantic category of the shown stimuli with high classification accuracy  using EEG signals. We proposed an attentional-driven convolutional neural network, specifically over occipital brain areas, due to its importance when processing visual information. Additionally, we compared our results with conventional approaches. Our findings lead to better classification accuracy and showed that it is possible to classify EEG signals during the representation of visual stimuli into different semantic categories with high accuracy. This proves its feasibility to be used in an intuitive BMI and can be further applied to decode visual imagery paradigms.

\section{Materials and Methods}

\subsection{Dataset Description}
We used the dataset published by Kaneshiro et al.~\cite{kaneshiro}. EEG signals were measured from ten healthy subjects (3 females, 21-57 years old) with normal color and normal or corrected-to-normal vision.

Subjects were shown 12 photographs of the six following semantic categories: human body (HB), human face (HF), animal body (AB), animal face (AF), fruits or vegetables (FV), and inanimated objects (IO), for a total of 72 images set against a mid-gray background. The experimental paradigm consisted of the presentation of an image during 500 msec and a black-gray screen during 750 msec. Subjects performed a total of 5,184 trials divided into 2 sessions. Each session consisted of 3 blocks, containing 864 trials each (each image was shown 12 times randomly) with short breaks every 36 trials. Therefore, each image was shown 72 times to each subject. The experimental paradigm was implemented using Neurobehavioral System Presentation software. EEG data were recorded using unshielded 128-channel EGI HCGSN 110 nets with a sampling rate of 1,000 Hz. 

EEG signals were preprocessed using a high-pass fourth-order Butterworth filter to remove frequency below 1 Hz and a low-pass eight-order Chebyshev Type I filter to remove frequency above 25 Hz. Additionally, data was subsampled to 62.5 Hz and the Infomax independent component analysis (ICA) algorithm was performed to remove ocular artifact and 124 channels were retained to further analyze. 

\begin{table}[!b]
\caption{CNN Block Description}
\label{CNN}
\centering
\renewcommand{\arraystretch}{1.0}
\begin{tabular}{cccc}
 \hline
Layer  & Operation         & Kernel Size & Feature Map / Neuron \\  \hline
\multicolumn{1}{c}{\multirow{2}{*}{1}} & Convolution         & (1, 5)  & 20  \\
\multicolumn{1}{c}{}                   & Batch Normalization &         &     \\
\multirow{2}{*}{2}                     & Convolution         & (124, 1) & 20  \\
                                       & Batch Normalization &         &     \\
\multirow{3}{*}{3}                     & Dropout (p = 0.5)     &         &     \\
                                       & Convolution         & (1, 5)   & 40  \\
                                       & Batch Normalization &         &     \\
\multirow{3}{*}{4}                     & Dropout (p = 0.5)     &         &     \\
                                       & Convolution         & (1, 10)  & 100 \\
                                       & Batch Normalization &         &     \\
\multirow{3}{*}{5}                     & Dropout (p = 0.5)     &         &     \\
                                       & Convolution         & (1, 10)  & 200 \\
                                       & Batch Normalization &         &         \\
Output & Fully-Connected & -           & 6 or 72-class                    \\\hline
\end{tabular}
\end{table}

\begin{figure*}[!t]
\centering
\includegraphics[width=0.85\textwidth]{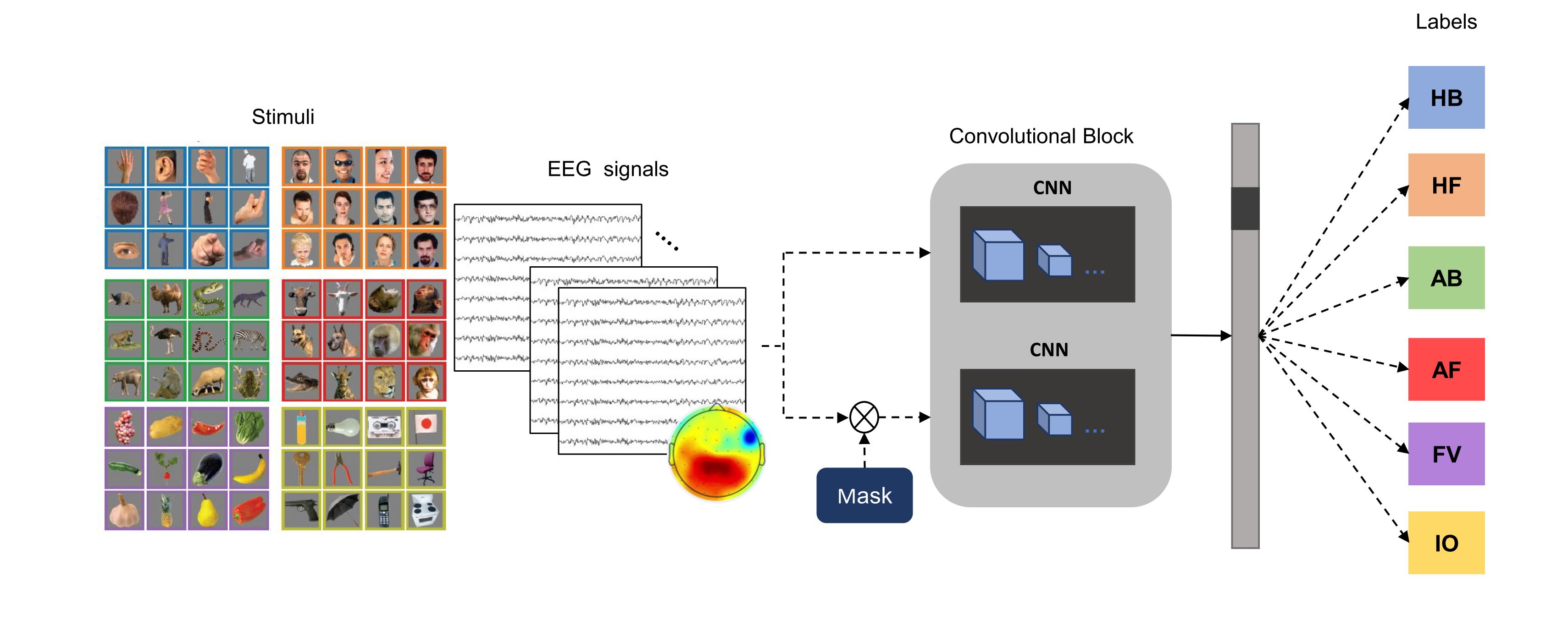}
\caption{Proposed framework. In the left, the shown stimuli are presented separated by their respective semantic category. Blue section correspond to human body (HB), orange to human face (HF), green to animal body (AB), red to animal face (AF), purple to fruit and vegetable (FV), and yellow to inanimated objects (IO). A mask over the occipital area is applied to the input (124 $\times$ 32). The output of the convolutional block is concatenated to a fully connected layer with softmax activation for classification. Classification is performed in semantic category level (6-class) and exemplar level (72-class). }
\label{arch}
\end{figure*}
\subsection{Proposed Method}
We proposed an attention driven CNN to classify the shown stimuli in different semantic categories.  The model receives as input the preprocessed data with the size of 124 $\times$ 32 (channels $\times$ time points). The model has two blocks of CNN with 5 layers (see Table~\ref{CNN}) with rectified linear unit (ReLU) as activation function, one block receives as input the preprocessed data and the other block receives as input the signal after applying a mask over occipital electrodes; this was done to drive the model attention to those areas considering that visual information is first processed in the occipital area of the brain \cite{goodale}. Additionally, dropout (\textit{p} = 0.5) and batch normalization were used to avoid overfitting of the model \cite{Ioffe, Schirr}. The output of the two CNN blocks is concatenated in a fully connected layer with a softmax activation. Loss is minimized using cross-entropy (equation 1) and parameters are optimized via Adam method. 
\begin{align} 
\label{eq:multinom}
  H_p(q)= -\sum_{c=1}^C y_c\times\ log(p(y_c))
\end{align}
where \textit{C} indicates number of classes, \textit{y} represent a binary indicator of class label [0, 1], and \textit{p} means predicted probability of belonging to that class. We used a grid search to tune the parameters and decided to use 25 epochs to train the model and 0.005 as the learning rate. Figure~\ref{arch} shows the proposed architecture.
We applied 10-fold cross-validation. Data of each subject was divided into 10 groups, 9 of the 10 groups which represent 90\% of the data was used as the training set and the remaining one as the test set. This was done until all the groups were used as test set. We averaged the accuracy of the test set and reported in the results section. As the performance measure, precision, recall, and F1-measure were calculated. 

\subsection{Other Methods}
We implement Shallow ConvNet proposed in Schirrmeister et al.~\cite{Schirr}, long short term memory network (LSTM), and LSTM + CNN. Shallow ConvNet consists of two CNN layers and a fully connected layer with a softmax activation, this architecture has successfully applied to EEG signals for the classification of MI paradigms \cite{Schirr}. LSTM model consists of 2 layers with 100 hidden units architecture, which has also been applied to classify EEG signals when classifying sleep stages and others~\cite{Hart}. Models that combine LSTM and CNN networks have been used before for studying EEG signals~\cite{Zhang}. We implement an LSTM + CNN model that consists of the application of an LSTM over the input signal follow by a CNN network, with a fully connected layer with softmax activation as the final layer. Finally, we also perform classification using the proposed CNN block without using the mask.

We also applied 10-fold cross-validation to evaluate the models, accuracies over test folds were averaged and reported in the results section.

\subsection{Statistical Analysis}
We used statistical analysis and performed a multiple comparison between the accuracy of the proposed and other methods. Paired t-test was applied and the significant level was set at \textit{p} = 0.05.

\
\section{Results}
\subsection{Classification using Proposed Model}
\begin{table}[!b]
\caption{Classification Accuracy using Proposed Method across All Subject (10-fold average $\pm$ standard deviation)}
\label{classification}
\centering
\renewcommand{\arraystretch}{1.2}
\begin{tabular}{cccc} 
\hline
\multicolumn{1}{c}{\multirow{1}{*}{Subject}} & 6-class accuracy (\%)      & 72-class accuracy (\%)       \\  \hline
Sub01                                         & 47.55 $\pm$ 1.75 & 26.45 $\pm$ 2.25  \\
Sub02                                         & 43.70 $\pm$ 2.50 & 18.03 $\pm$ 1.79  \\
Sub03                                         & 51.27 $\pm$ 2.25 & 28.55 $\pm$ 2.37  \\
Sub04                                         & 45.27 $\pm$ 1.47 & 14.34 $\pm$ 1.49  \\
Sub05                                         & 58.40 $\pm$ 1.37 & 39.19 $\pm$ 2.70  \\
Sub06                                         & 58.50 $\pm$ 2.40 & 42.81 $\pm$ 2.22  \\
Sub07                                         & 55.43 $\pm$ 1.97 & 29.16 $\pm$ 2.56  \\
Sub08                                         & 40.08 $\pm$ 1.42 & 13.79 $\pm$ 1.74  \\
Sub09                                         & 45.22 $\pm$ 3.73 & 16.27 $\pm$ 1.77  \\
Sub10                                         & 58.32 $\pm$ 2.37 & 38.95 $\pm$ 2.87  \\  \hline
\textbf{Average}                                     & \textbf{50.37 $\pm$ 6.56 } & \textbf{ 26.75 $\pm$ 10.38}  \\ \hline
\end{tabular}%
\end{table} 

\begin{table*}[!htbp]
\caption{ Recall and F1-measure values per Semantic Category (6-class) across All Subjects}
\label{class_eval}
\centering
\renewcommand{\arraystretch}{1.2}
\begin{tabular}{ccccccccccccc}
\hline
\multirow{2}{*}{Subject} & \multicolumn{2}{c}{HB} & \multicolumn{2}{c}{HF} & \multicolumn{2}{c}{AB} & \multicolumn{2}{c}{AF} & \multicolumn{2}{c}{FV}  & \multicolumn{2}{c}{IO} \\
                          & Recall   & F1  & Recall   & F1  & Recall   & F1  & Recall   & F1  & Recall & F1     & Recall   & F1  \\ \hline
Sub01                     & 0.495    & 0.406       & 0.589    & 0.651       & 0.445    & 0.443       & 0.455    & 0.494       & 0.345  & 0.309          & 0.492    & 0.515       \\
Sub02                     & 0.404    & 0.365       & 0.578    & 0.648       & 0.406    & 0.426       & 0.390    & 0.453       & 0.347  & 0.310          & 0.464    & 0.499       \\
Sub03                     & 0.504    & 0.490       & 0.713    & 0.757       & 0.553    & 0.564       & 0.434    & 0.481       & 0.480  & 0.354          & 0.446    & 0.488       \\
Sub04                     & 0.371    & 0.366       & 0.656    & 0.710       & 0.428    & 0.454       & 0.501    & 0.520       & 0.307  & 0.293 & 0.412    & 0.467       \\
Sub05                     & 0.529    & 0.498       & 0.724    & 0.783       & 0.611    & 0.609       & 0.512    & 0.526       & 0.521  & 0.364          & 0.591    & 0.564       \\
Sub06                     & 0.573    & 0.599       & 0.732    & 0.783       & 0.609    & 0.612       & 0.556    & 0.582       & 0.509  & 0.483          & 0.532    & 0.450       \\
Sub07 & 0.496    & 0.458       & 0.681    & 0.751       & 0.550    & 0.575       & 0.549    & 0.544       & 0.504  & 0.360          & 0.520    & 0.530       \\
Sub08 & 0.340    & 0.341       & 0.593    & 0.620       & 0.433    & 0.441       & 0.390    & 0.374       & 0.312  & 0.290 & 0.324    & 0.336       \\
Sub09 & 0.388    & 0.391       & 0.736    & 0.713       & 0.415    & 0.373       & 0.571    & 0.555       & 0.327  & 0.302          & 0.348    & 0.423       \\
Sub10 & 0.489    & 0.416       & 0.742    & 0.795       & 0.539    & 0.593       & 0.676    & 0.600       & 0.517  & 0.363          & 0.478    & 0.507      \\ \hline
\textbf{Average}  & \textbf{0.459 } & \textbf{ 0.433 } & \textbf{ 0.674 } & \textbf{ 0.721} & \textbf{ 0.499 }& \textbf{ 0.509 }& \textbf{ 0.503}& \textbf{ 0.513 }& \textbf{ 0.417 }& \textbf{ 0.343 }& \textbf{ 0.461 }& \textbf{ 0.478 }\\ \hline
\textbf{std.}  & \textbf{0.073} & \textbf{ 0.074 } & \textbf{ 0.062 } & \textbf{ 0.059} & \textbf{ 0.077 }& \textbf{ 0.085 }& \textbf{ 0.085}& \textbf{ 0.062 }& \textbf{ 0.090 }& \textbf{ 0.054 }& \textbf{ 0.078 }& \textbf{ 0.060 }\\ \hline
\end{tabular}
\end{table*}

\begin{figure}[!htbp]
\centering
\includegraphics[width=3.4in]{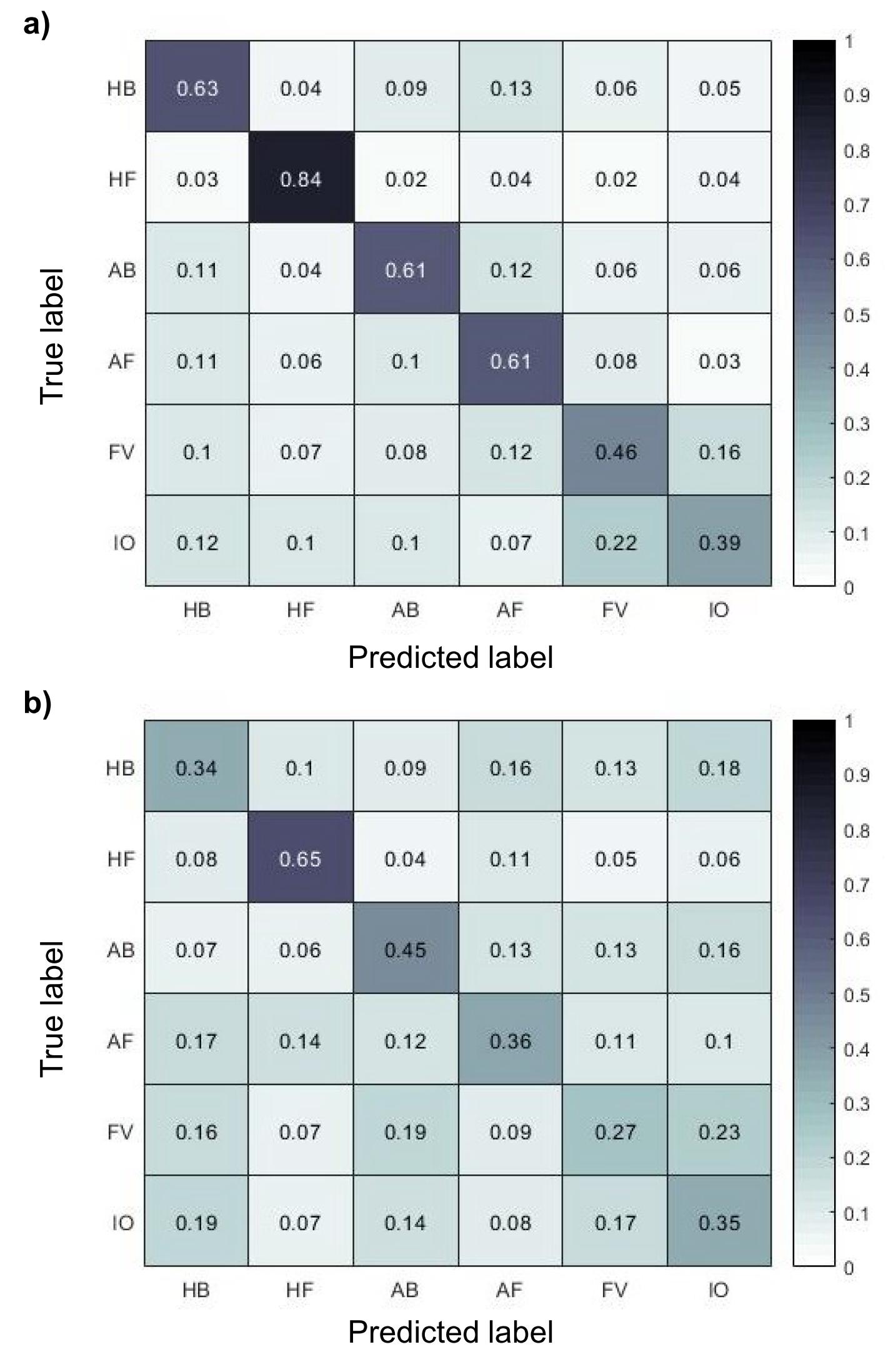}
\caption{Confusion matrix of each class from representative subjects. (a) Sub06 (the highest performance) and (b) Sub08 (the lowest performance).}
 \label{confmatrix}
\end{figure}

Table~\ref{classification} shows the classification accuracy of the proposed model across all subjects. When classifying stimulus depending on the semantic category they belong (6-class), the proposed model achieved an accuracy of 50.37 $\pm$ 6.56\%. On the other hand, when classifying the stimulus as the image presented (72-class), the proposed model achieved an accuracy of 26.75 $\pm$ 10.38\%. Sub06 had the highest performance (58.50 $\pm$ 2.40\% and 42.81 $\pm$ 2.22\% for 6-class and 72-class, respectively) while Sub08 had the lowest (40.08 $\pm$ 1.42\% and 13.79 $\pm$ 1.74\% for 6-class and 72-class, respectively). The obtained classification accuracies are significantly higher than the chance level (6-class: 16.67\% and  72-class: 1.38\%). 
Table ~\ref{class_eval} shows recall and F1-measure values (values obtained per fold are averaged and presented) per class (6-class) across all subjects. It is observed that HF is more distinguishable than other classes across all subjects. Meanwhile, values of recall and F1-measure are lowest for FV class across all subjects. Obtained values show that our model can distinguish between semantic categories. 

Figure~\ref{confmatrix} shows the sum of the confusion matrix obtained in each fold and normalized in Sub06 and Sub08 for 6-class. The model mostly confuses FV class with IO class, while HF class is the most separable class for both cases. For Sub08, the model also confuses HB and IO class, additionally, the model also confuses FV and AB class.

\begin{figure*}[!t]
\centering
\includegraphics[width=0.9\textwidth]{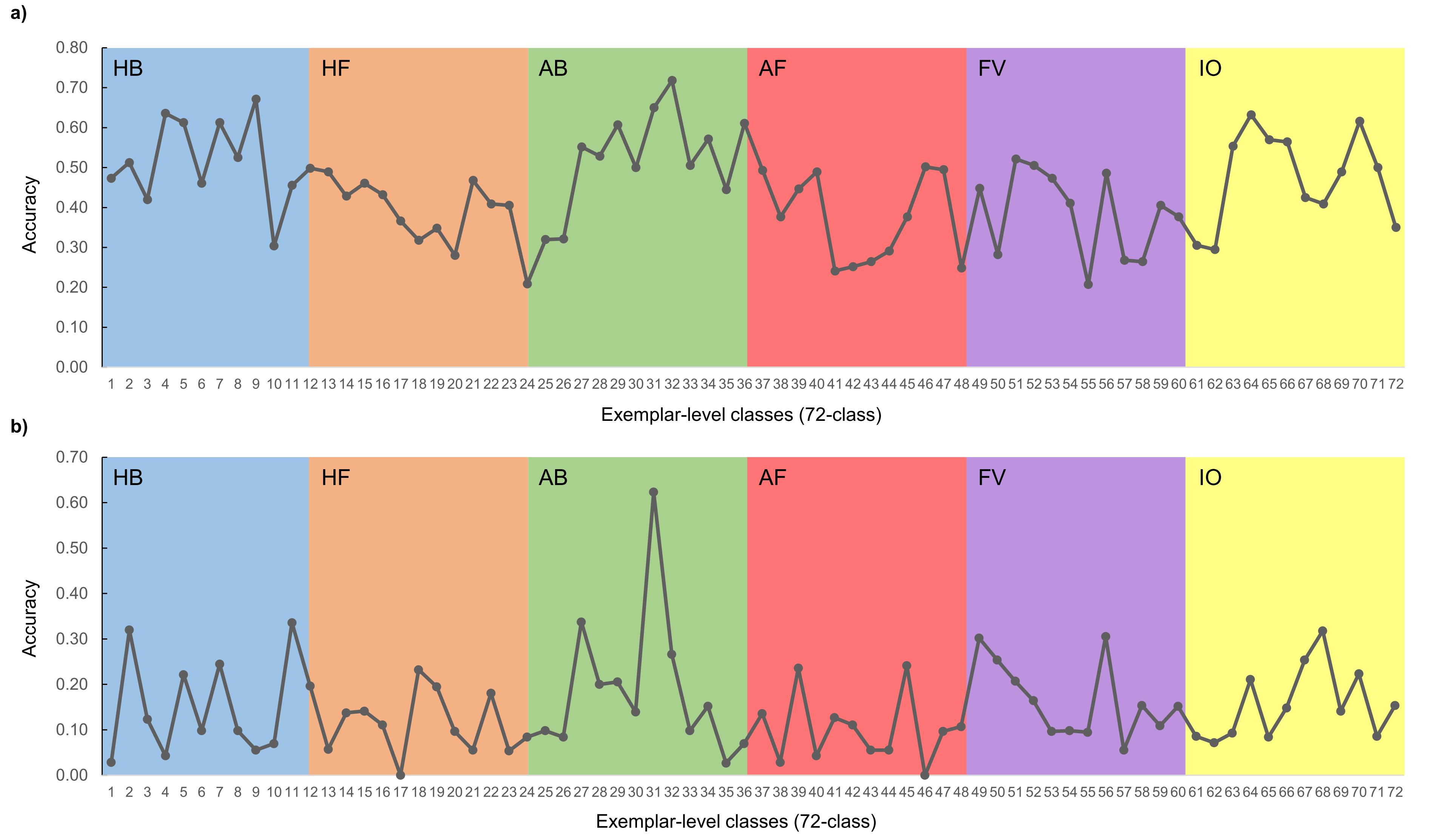}
\caption{Classification results of each exemplar from representative subjects (a) Sub06 and (b) Sub08. Blue section correspond to human body (HB), orange to human face (HF), green to animal body (AB), red to animal face (AF), purple to fruit and vegetable (FV), and yellow to inanimated objects (IO).}
 \label{acc_class}
\end{figure*} 
Figure~\ref{acc_class} depicts the classification results for the exemplar-level classes (72-class) for Sub06 and Sub08. It is observed that classification accuracy varies from class to class even inside the same semantic category group. Specifically, the exemplar classes of the AB category are more distinguishable than other categories. 

\subsection{Comparison of Other Methods}
We compared the proposed model with other methods. Table~\ref{comparison} shows the classification accuracy using different methods across subjects and the results of the statistical analysis when compared different methods to the proposed one. The lowest classification accuracy was obtained when applying LDA as a classification method 40.68 $\pm$ 5.54\% and 14.46 $\pm$ 6.43\% for 6-class and 72-class, respectively. In addition, 6-class performance is 44.77 $\pm$ 6.40\%, 46.18 $\pm$ 6.79\% and  49.04 $\pm$ 6.99\% and 72-class performance is 15.39 $\pm$ 6.01\%, 23.23 $\pm$ 10.48\% and 23.72 $\pm$ 10.95\% for LSTM, LSTM + CNN,  and Shallow ConvNet, respectively. When using CNN without the mask, an accuracy of 50.00 $\pm$ 6.61\% and 25.93 $\pm$ 10.67\% was obtained for 6-class and 72-class, respectively. Statistical analysis exhibits significant differences between the proposed and all other conventional methods (\textit{p} $<$ 0.05).
\begin{table}[!t]
\caption{Comparison of Classification Performance using Different Methods and Statistics between Proposed and other Methods (*\textit{p}-value $<$ 0.05, **\textit{p}-value $<$ 0.01)}
\label{comparison}
\centering
\renewcommand{\arraystretch}{1.5}
\begin{tabular}{ccccccccccccc}
\hline
\multirow{1}{*}{Method} & \multicolumn{1}{c}{6-class accuracy (\%) }    & \multicolumn{1}{c}{72-class accuracy (\%) }    \\ \hline
LDA [2]                    & 40.68 $\pm$ 5.54 **& 14.46 $\pm$ 6.43 **\\
Shallow~\cite{Schirr}          & 49.04 $\pm$ 6.99 * & 23.72 $\pm$ 10.95 **\\
LSTM                    & 44.77 $\pm$ 6.30 **& 15.39 $\pm$ 6.01 **\\
LSTM + CNN              & 46.18 $\pm$ 6.79 **& 23.23 $\pm$ 10.48 **\\
CNN              & 50.00 $\pm$ 6.61 * &  25.93 $\pm$ 10.67 *\\
Proposed              & 50.37 $\pm$ 6.56    & 26.75 $\pm$ 10.38    \\\hline
\end{tabular}%
\end{table}

\section{Discussion}
Our results show that it is possible to classify EEG signals during the presentation of a stimuli into semantic categories with higher accuracy than other methods. The obtained accuracy was 50.37 $\pm$ 6.56\% and 26.75 $\pm$ 10.38\% for 6-class and 72-class, respectively. We classified EEG signals using the CNN network and applying an attention mask to the input. The classification performance using attention was significantly higher than using only CNN. This can be attributed to the importance of occipital areas when processing visual information. Human face class is the most distinctive category; previous studies have explored brain responses to human faces, some of them concluded that there is a specific region of our brain dedicated to processing faces~\cite{haxby}. This can explain the high accuracy obtained for this semantic category. At the same time, fruit and vegetable category was the less distinctive category, which is in accordance with previous works~\cite{kaneshiro}. When classifying exemplar-level we could see that accuracies vary inside each category. Brain activation changes depending on the semantic category of the presented stimuli, however, certain semantic categories can cause similar activity, this confuses the model during classification, which can also explain the variability of exemplar-level classification inside each semantic category. This can be solved by increasing the spatial resolution of the data (e.g. use of more electrodes).

We additionally explore if using semantic category information over exemplar-level classification can increase the accuracy. We trained the model using a semantic category label and performed transfer learning. Weights of the CNN layers were kept and only weights of the last layer (fully connected layer) were updated, however, classification accuracy drop to 20.03 $\pm$ 7.13\%. This shows that even semantic categories can provide relevant information for classification, each of the exemplar-level classes carries specific information which increases the accuracy.

One limitation of this study is the number of trials when performing classification over exemplar-level class (72-class). Previous studies have shown that it is possible to classify EEG signals regardless of the small data quantity. Ten-fold cross-validation was performed to evaluate the model more generally overall data. Additionally, we used a predefined mask for all trials, however as mentioned before there are some specific areas of the brain in the occipital cortex to process certain visual information. Therefore, we decide to explore the applications of masks over different areas in order to improve classification accuracy. 

\section{Conclusion}
In conclusion, this study investigated the classification of presented images into semantic category and exemplar-level (6-class and 72-class) using different methods. Specifically, we evaluated the proposed CNN using and without using a mask over occipital regions, obtaining the highest classification accuracies when using it. This shows the importance of occipital regions when decoding visual objects. We decided to further explored classification methods for improving accuracy and apply our method to visual imagery paradigm. Even though visual perception and visual imagery share similar cortical representations, they have different flow in the visual pathways~\cite{dentico}, therefore additional research needs to be done regardless of visual imagery and the possibility of using a mask to improve classification. The use of this paradigm can provide a more intuitive and endogenous BMI for interpreting EEG signals and can improve the interaction with people that, due to various causes, are not able to communicate using conventional methods.As a result, its use can contribute to the development of a wide range of reliable BMI applications~\cite{kosmyna}.

\bibliography{IEEEexample.bib}{}
\bibliographystyle{IEEEtran}
\end{document}